\documentclass[aps,prl,superscriptaddress,floatfix,amsmath,amsfonts,twocolumn]{revtex4}  
\usepackage{enumitem}
\usepackage{amsmath}
\usepackage{graphicx}
\usepackage{epsfig}
\usepackage{bm}
\usepackage{bbold}
\usepackage{amssymb}
\usepackage{color}

\newcommand{\cE}{{\cal E}}

\begin{document}

\title{Adiabatic theory of one-dimensional curved polariton waveguides}

\author{D. A. Zezyulin\footnote{email: d.zezyulin@gmail.com}}

\affiliation{Department of Physics, ITMO University, Saint Petersburg 197101, Russia}

\author{I. A. Shelykh}

\affiliation{Science Institute, University of Iceland, Dunhagi 3, IS-107, Reykjavik, Iceland}
\affiliation{Department of Physics, ITMO University, Saint Petersburg 197101, Russia}

\date{\today}

\begin{abstract}

We construct a general theory of adiabatic propagation of spinor exciton-polaritons in waveguides of arbitrary shape, accounting for the effects of TE-TM splitting in linear polarizations and Zeeman splitting in circular polarizations. The developed theory is applied for the description of waveguides of periodically curved shape. We show  that in this geometry the periodic rotation of the effective in-plane magnetic field produced by TE-TM interaction results in a nontrivial band-gap structure, which can be additionally tuned by application of an external magnetic field. It is also demonstrated, that spin-dependent interactions between polaritons lead to the formation of stable gap solitons.


\end{abstract}

\maketitle

\textit{Introduction.}
Exciton-polaritons are composite half-light half-matter quasiparticles emerging in the regime of the strong coupling between a photonic mode of a planar semiconductor microcavity and an exciton in a quantum well  (QW) brought in resonance with it. They possess a set of remarkable properties, which allow polaritonic systems to serve as a convenient playground for study of collective nonlinear phenomena at elevated temperatures \cite{Ciuti2013}. From their photonic component polaritons get extremely small effective mass (about $10^{-5}$ of the mass of free electrons) and macroscopically large coherence length \cite{Ballarini2017}, while
the presence of an excitonic component enables  efficient polariton-polariton interactions \cite{Glazov2009,Vladimirova2010,Estrecho2019}  and leads to the sensitivity of the polariton systems to external electric \cite{Schneider2013,Suarez2020,Marin2022} and magnetic \cite{Solnyshkov2008,Walker2011,Krol2019} fields.

An important property of cavity polaritons is their spin (or pseudo-spin) \cite{ShelykhReview},
inherited from the spins of QW excitons and cavity photons.
Similar to photons, polaritons have two possible spin projections on the structure growth axis corresponding to the two opposite circular polarizations which can be mixed by effective magnetic fields of various origin. 
Real magnetic field applied along the structure growth axis and acting on the excitonic component splits in energy the polariton states with opposite circular polarizations, while TE-TM splitting of the photonic modes of a planar resonator couples these states to each other via a $k$-dependent term, thus playing a role of an effective spin-orbit interaction \cite{ShelykhReview}. 
Importantly, polariton-polariton interactions are also spin dependent, as they stem from the interactions of  excitonic components  which are dominated by the exchange term \cite{Ciuti1998}.
This leads to the fact that polaritons of the same circular polarization interact orders of magnitude stronger than polaritons with opposite circular polarizations \cite{Glazov2009}.

Remarkable tunability of cavity polaritons allows to engineer their spatial confinement in a variety of experimental geometries, ranging from individual micropillars \cite{Bajoni2008,Ctistis2010,Ferrier2011,Real2021} to   systems of several coupled pillars forming so-called polariton molecules \cite{Galbiati2012,Sala2015} or periodically arranged arrays of the pillars forming polariton superlattices \cite{Milicevic2017,Suchomel2018,Whittaker2018,Whittaker2021,Kuriakose2022}. Realization of quasi one-dimensional (1D) geometries, where the motion of the polaritons is restricted to individual waveguides \cite{Sich2018,Suarez2020}, rings \cite{Lukoshkin2018,Mukherjee2019,Sedov2021} or systems of coupled waveguides \cite{Winkler2017,Belerin2021}, represents particular interest from the point of view of the applications of polaritonics, as they can form basis for classical \cite{Liew2010,Shelykh2011,Chen2022}
 and quantum \cite{Xue2021,Nigro2022} polaritonic circuits. 

 Current state of technology allows routine production of quasi 1D polariton waveguides of arbitrary shape, including ones with periodically modulated curvature. Creation of the general theory of the polariton propagation in these structures, which includes polarization dynamics and polariton-polariton interactions, is the goal of the present Letter.

\begin{figure}
   \begin{center}
		\includegraphics[width=0.99\columnwidth]{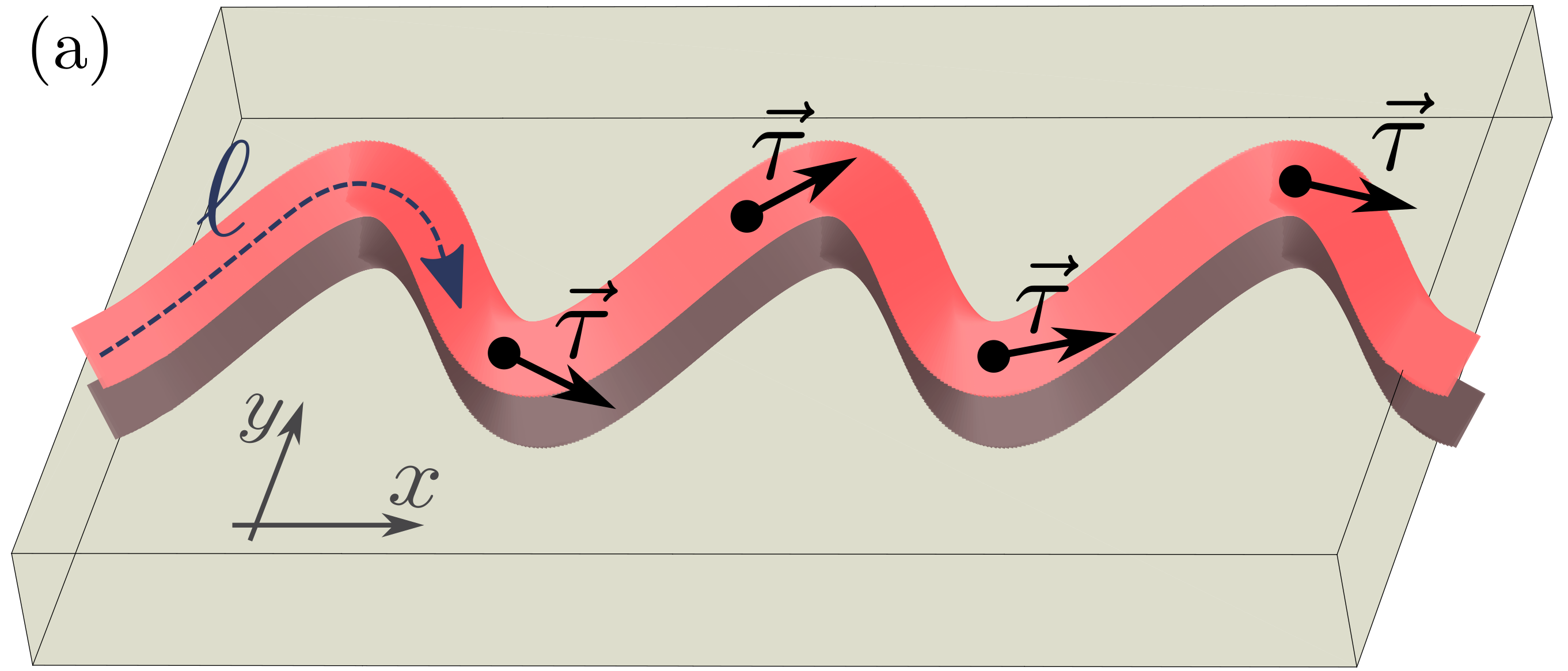}\\[4mm]%
  \includegraphics[width=\columnwidth]{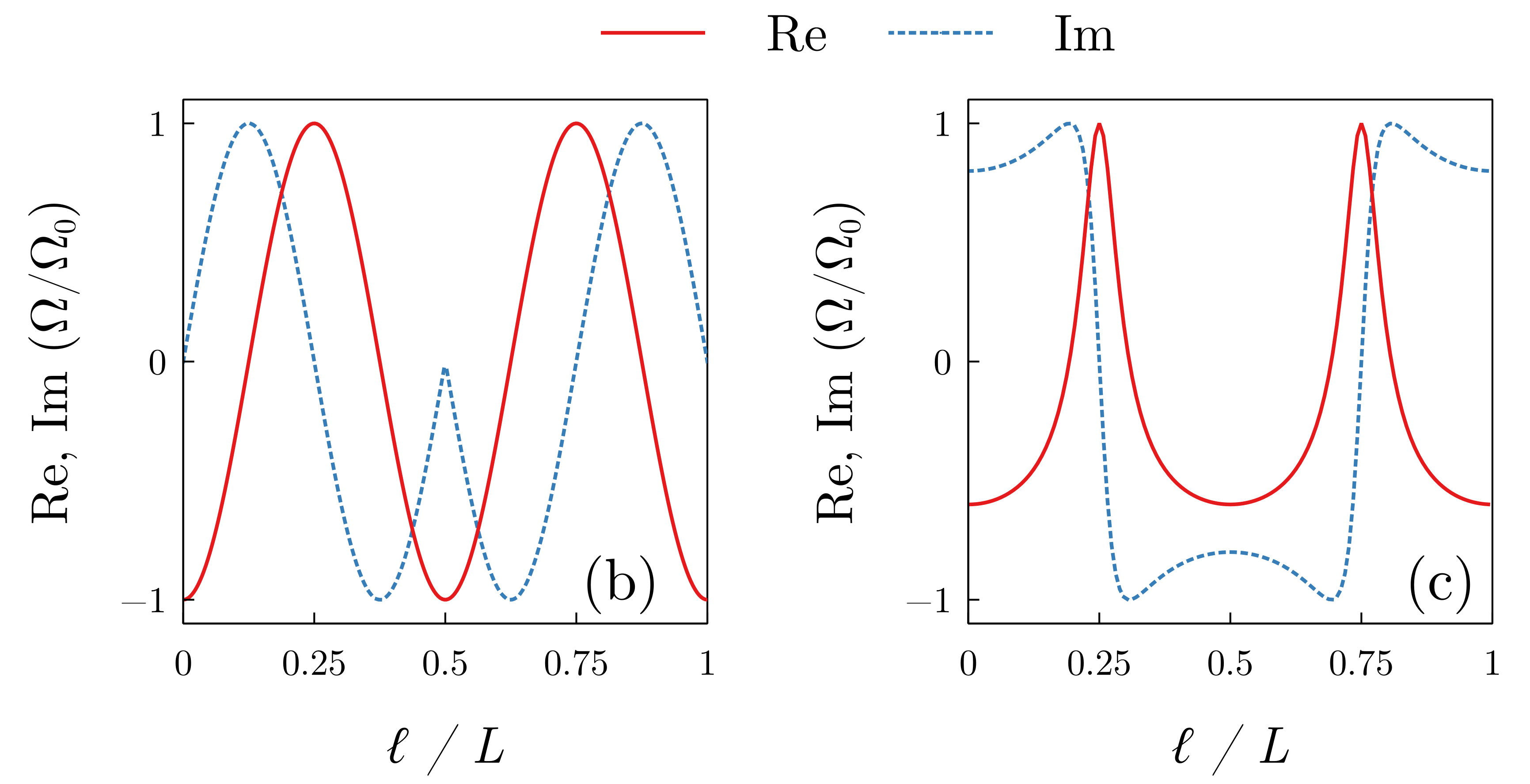}%
	\end{center}
    \caption{(a) Schematic representation of the considered geometry of a 1D polariton waveguide etched in planar semiconductor microcavity. The arc length $\ell$ measures the distance along the waveguide. Direction of the in-plane tangential unit vector $\vec{\tau} = (\tau_x, \tau_y)$ changes along the waveguide and leads to emergence of an effective space-dependent field for the spinor polariton wavefunction. (b,c) Real and imaginary parts of the $L$-periodic effective potentials $\Omega(\ell)$ for a waveguide composed of a chain of touching halfcircles (b) and a sine-shaped waveguide (c). }
    \label{fig:idea}
\end{figure}

 \textit{The model.} The presence of the in-plane spatial confinement results in  the strong nonequivalency of the states polarized normally and tangentially to a waveguide, which leads to the appearance of a local effective magnetic field, acting on a polariton pseudospin and directed tangentially to  the waveguide. Although one can safely assume  that in the case of a narrow waveguide of a constant width the absolute value of this field remains constant (see Supplementary material \cite{SupplMatRef} for further details), its direction changes along the curved waveguide, and, as we demonstrate below, this has crucial effect on polariton dynamics.

 Let us suppose that the shape of a waveguide in $(x, y)$-plane is given parametrically as $x=x(\xi), y=y(\xi)$. The components of the effective magnetic field $\Omega_{x,y}$ produced by TE-TM interaction are proportional to the components of the unit vector tangential to a waveguide $\tau_{x,y}$ and thus read
 \begin{eqnarray}
\Omega_{x}=\Omega_0\tau_{x}=\frac{\Omega_0x'(\xi)}{\sqrt{x'(\xi)^2+y'(\xi)^2}},\\
\Omega_{y}=\Omega_0\tau_{y}=\frac{\Omega_0y'(\xi)}{\sqrt{x'(\xi)^2+y'(\xi)^2}},
 \end{eqnarray}
where primes correspond to derivatives, and 
\begin{equation}
\Omega_0\approx\frac{\hbar^2}{4d^2}\left(\frac{1}{m_l}-\frac{1}{m_t}\right).
\end{equation}
In the above equation, $m_l$ and $m_t$ stand for the effective longitudinal and transverse masses of 2D polaritons, and $d$ is an effective width of a polariton channel \cite{Shelykh2018}. As it was already mentioned, the presence of the field $\boldsymbol\Omega$ splits in energy the modes polarized normally and tangentially to a waveguide. Additional splitting in circular polarizations, denoted by $\Delta_z$, can be induced by application of an external magnetic field perpendicular to a cavity interface. 

Let us introduce the coordinate $\ell$ along the waveguide, 
$\ell = \int_0^\xi \sqrt{x'(\eta)^2 + y'(\eta)^2} d\eta
$. In the adiabatic approximation, the effective  1D Hamiltonian governing the dynamics of the spinor wavefunction of polaritons can be then represented in the following form (see Supplementary material \cite{SupplMatRef} for corresponding derivation):
\begin{equation}
\label{eq:H}
   \hat{H} = \left(\begin{array}{cc}
        \displaystyle -\frac{\hbar^2}{2m_{eff}}\frac{d^2\ }{d\ell^2} + \frac{\Delta_z}{2}&  \Omega_-  \\[3mm]
        \Omega_+& \displaystyle  -\frac{\hbar^2}{2m_{eff}}\frac{d^2\ }{d\ell^2} -\frac{\Delta_z}{2}
    \end{array}\right),
\end{equation}
where  
\begin{equation}
\Omega_\pm=\Omega(\ell) =  \Omega_0(\tau_x \pm i\tau_y)^2,     
\end{equation}
and $m_{eff}$ is the effective  mass. 

The physical meaning of the above Hamiltonian is pretty clear: it describes a motion of a one-dimensional spinor particle affected by a constant $z$-directed magnetic field and in-plane magnetic field whose direction changes along the way, being always tangential to the waveguide.

In what follows, we will work with the effective Hamiltonian rewritten in the dimensionless form. To this end, we introduce the unit length $\lambda_0$ and the unit energy $\varepsilon_0 \equiv \hbar^2/(2m_{eff} \lambda_0^2)$, and then rescale the variables of (\ref{eq:H}) as $\ell \to \lambda_0 \ell$ and  $\Delta_z \to \varepsilon_0 \Delta_z$. Additionally, we rescale time as $t \to (\hbar/\varepsilon_0) t$. Assuming, for instance, that the   unit length $\lambda_0$ corresponds to 5~$\mu$m and $m_{eff}$ is about $10^{-5}$ of the free electron mass, we obtain that the  unit  energy $\varepsilon_0$ is about 0.2~meV, and the time unit $\hbar/\varepsilon_0$ is equivalent to few picoseconds. Supplementing the obtained dimensionless Hamiltonian with the interaction terms \cite{Flayac2010}, we obtain the following nonlinear evolution problem that governs the dynamics of the spinor wavefunction $(\Psi_1, \Psi_2)$:
\begin{eqnarray}
\label{eq:nonlin1}
     i\frac{\partial \Psi_1}{\partial t} = -\frac{\partial^2 \Psi_1}{\partial \ell^2} + \frac{\Delta_z}{2} \Psi_1  + \Omega_-(\ell) \Psi_2 \nonumber\\[1.0mm]+ (|\Psi_1|^2 + \sigma |\Psi_2|^2)\Psi_1,\\[2mm]
\label{eq:nonlin2}     
    i\frac{\partial \Psi_2}{\partial t} = -\frac{\partial^2 \Psi_2}{\partial \ell^2}  - \frac{\Delta_z}{2} \Psi_2  + \Omega_+(\ell) \Psi_1 \nonumber\\[1.0mm] + (|\Psi_2|^2 + \sigma |\Psi_1|^2)\Psi_2.
\end{eqnarray}
Small negative coefficient $\sigma$ takes into account weak attraction between   polaritons of opposite polarizations (in our numerical calculations the value $\sigma=-0.05$ was used).


\begin{figure*}
   \begin{center}
        \includegraphics[width=1.\textwidth]{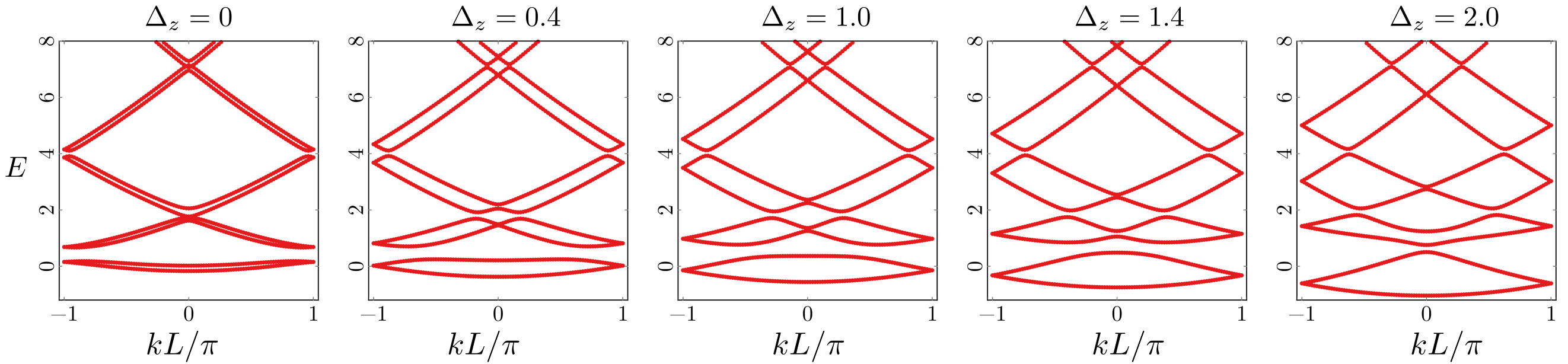}%
    \caption{Transformation of the band-gap structure for  the sine-shaped waveguide under the fixed   TE-TM splitting coefficient $\Omega_0 = 0.45$  and increasing strength of the external magnetic field $\Delta_z$. Here the Bloch quasimomentum $k$ varies within the reduced Brillouin zone $[-\pi/L, \pi/L)$, where $L$ is the spatial period of the structure. The periodic curvature results in a nontrivial band-gap structure. Finite bandgaps are present even in the absence of the external magnetic field ($\Delta_z = 0)$.  The increase of $\Delta_z$ leads  to the anticrossings of the bands touching at $k=0$ and related shift of the band minima and maxima to $k\neq0$.}
    \label{fig:bands}
    \end{center}    
\end{figure*}

\textit{Examples: The chain of halfcircles and the sine-shaped waveguide.} 
In what follows, we focus on the situation when the shape of the curved waveguide can be described by function $y(x)$,  see Fig.~\ref{fig:idea}(a) for a schematics of the assumed geometry.  Then the effective field, as a function of the arc length $\ell$, can be computed as $\Omega_\pm(\ell) =  \Omega_0\exp\{ \pm 2 i \arctan(dy/dx)\}$, where the derivative $dy/dx$ should be expressed as a function of $\ell$. In our further consideration we focus on the case of periodically curved waveguides.

As a first analytically tractable example we consider the situation when the waveguide is composed of a periodic chain of touching halfcircles of a radius $R$. In terms of coordinates $x$ and $y$, the unit cell of the  resulting periodic structure is given as $y(x) = \sqrt{R^2 - (x-R)^2}$ for $x\in [0, 2R]$ (the upper halfcircle) and $y(x) = -\sqrt{R^2 - (x-3R)^2}$ for $x\in [2R, 4R]$  (the lower halfcircle). In terms of the arc length $\ell$, the   unit cell
corresponds to the  interval $\ell \in [0, L]$ where $L= 2\pi R$ is the period of the structure.  The first halfperiod   $\ell \in [0, \pi R]$ corresponds to the first halfcircle, where $x(\ell)  = R[1-\cos (\ell/ R)]$  and $y(\ell) = R\sin (\ell /R)$, and the 
 second halfperiod  $\ell \in [\pi R, 2\pi R]$ corresponds to the  second  halfcircle, where we have parametrization $x(\ell)  = R[3+\cos (\ell/ R)]$  and $y(\ell) = R\sin (\ell /R)$, and the rest of waveguide is obtained by the periodic repetition of the unit cell. Performing straightforward calculations, we obtain that within the unit cell the resulting potential reads  $\Omega_\pm(\ell) = -\Omega_0 \exp\{\mp 2i\ell\, \textrm{sign}\,(\pi R-\ell)/R\}$.  The shape of the resulting dependency is illustrated in Fig.~\ref{fig:idea}(b). While the obtained dependence is rather simple, its imaginary part is not a smooth function: it has a cusp exactly at the center of the unit cell  $\ell=\pi R$, where the two halfcircles touch.

 As a second example, which results in a smooth periodic potential (which is therefore   better suited for the numerical analysis), we consider a sine-shaped waveguide  $y(x) = V_0 \sin x $.  Then the   arc length along the waveguide is given by the incomplete elliptic integral of  the second kind \cite{Olver10}: $\ell(x) = \sqrt{1+V_0^2}\, \cE(\sin x, m)$, where $m =  V_0^2/(1+V_0^2)$.
To the best of our knowledge, there is neither a commonly used special function nor a closed-form expression that allows to invert the incomplete elliptic integral of the second kind, i.e., to express $x$ and $y$ through $\ell$ in our case. In the meantime, there exists a simple  iterative numerical procedure  for  inversion of the incomplete elliptic integral of the second kind \cite{Boyd2012}. Using this procedure, one can easily obtain the dependence $\Omega(\ell)$, see Fig.~\ref{fig:idea}(c) for a representative example. The resulting 1D Hamiltonian $\hat{H}$ defined by (\ref{eq:H}) becomes effectively periodic with the spatial period in $\ell$ given as $L = 4\cE(m)$, where $\cE(m)$ is the complete elliptic integral of the second kind.

\textit{Band structure.} Periodic nature of the resulting system suggests to look at the   band structure which  can be presented in the form  of the dependencies of the energy $E$ versus Bloch quasimomentum $k$, 
which, without loss of generality, can be assumed to belong to the   Brillouin zone $[-\pi/L, \pi/L)$, where $L$ is the period. For sinusoidal waveguide the result computed for system (\ref{eq:nonlin1})--(\ref{eq:nonlin2}) with omitted nonlinear terms $(|\Psi_{1,2}|^2 + \sigma |\Psi_{2,1}|^2)\Psi_{1,2}$ is shown in Fig.~\ref{fig:bands}. We have focused on the transformation of the spectral structure subject the the increase of the external magnetic field, which is characterized by the Zeeman splitting coefficient $\Delta_z$. As one can see, the periodic curvature of a waveguide results in a nontrivial band-gap structure as the effective periodic potential $\Omega(\ell)$ opens finite gaps even in the absence of the external magnetic field ($\Delta_z = 0)$.  The increase of $\Delta_z$ leads to a transformation of the band-gap structure. In particular, it leads to the anticrossing of the bands touching at $k=0$ and related shift of the band minima and maxima to $k\neq0$.  Dispersion curves having two degenerate extrema at $k=\pm k_0 \ne 0$ can be, in particular, relevant for the observation of the so-called stripe phase characterized by spinor wavefunctions carrying a more complex internal structure, see e.g. \cite{Wang2010,Ho2011,Li2012,Achilleos2013,Kartashov2013} and \cite{Zezyulin2020} for   discussion of   stripe phase and stripe solitons in spin-orbit coupled atomic and polariton condensates, respectively.

\begin{figure} 
   \begin{center}
		\includegraphics[width=\columnwidth]{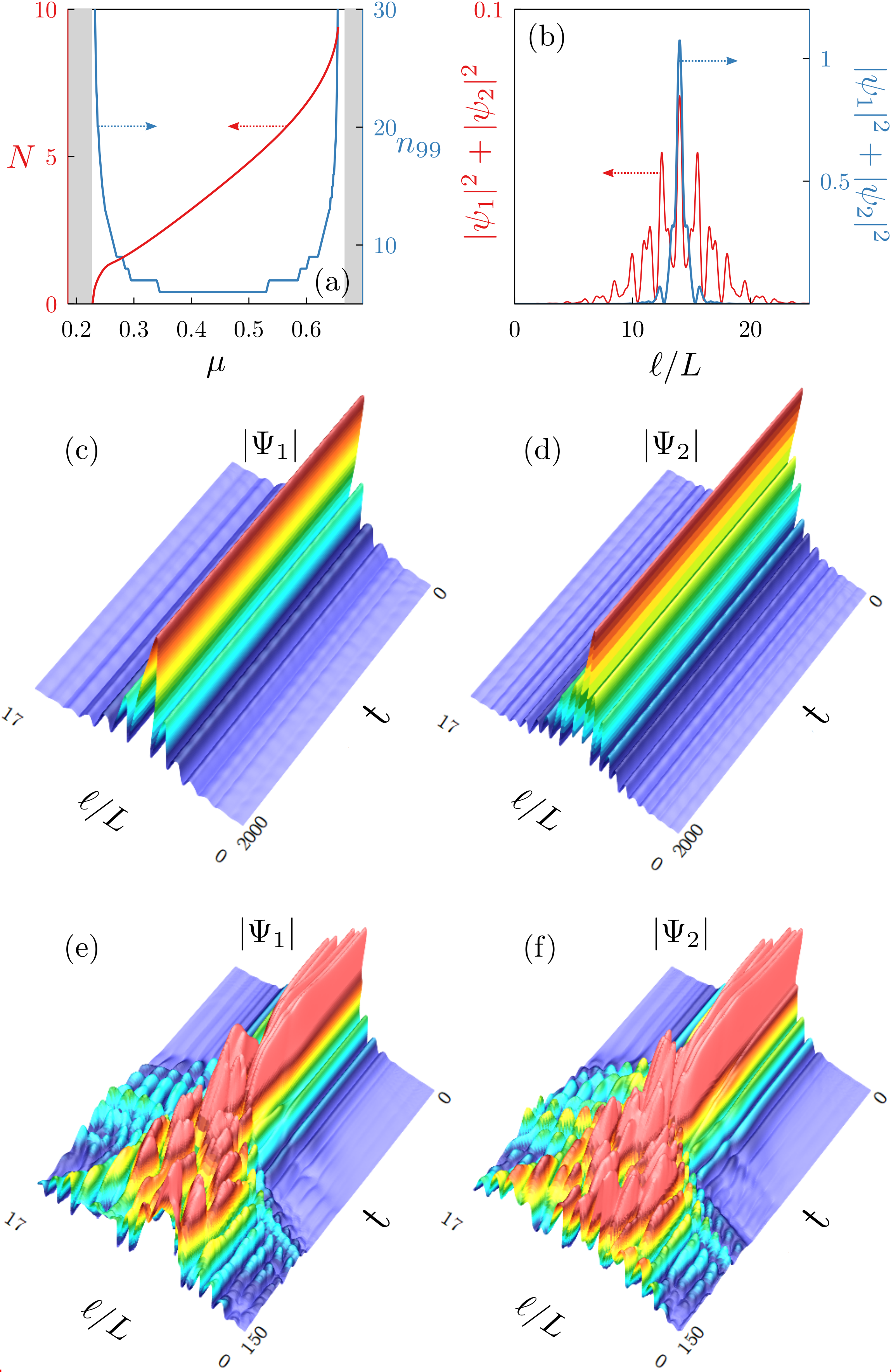}%
	\end{center}
    \caption{(a) Gap solitons norm $N$ and the localization measure $n_{99}$  as functions of chemical potential $\mu$ for a family of fundamental gap solitons in the first finite gap. Here the coefficient of TE-TM splitting $\Omega_0=0.4$ and amplitude of the Zeeman splitting $\Delta_z=0.3$. Shaded regions correspond to the values of $\mu$ that belong to   spectral bands. (b) Example of a  broad soliton near the left edge of the gap (specifically, at  $\mu=0.24$) and a strongly localized soliton in the center of the gap at $\mu=0.5$.  (c,d) Stable dynamics of the gap soliton with  chemical potential $\mu=0.29$. Initial conditions correspond to the stationary wavefunctions perturbed with a     random noise whose amplitude is about $2\%$ of the soliton's amplitude.  (e,f) Example of unstable evolution of a gap soliton of  more complex shape corresponding to $\Omega_0=0.4$, $\mu = 0.4$, and $\Delta_z = 0.009$.}
    \label{fig:family}
\end{figure}

\textit{Gap solitons.} The presence of finite gaps in the band-gap structure suggests that when the repulsive interactions between the polaritons of the same circular polarization are taken into account, the waveguide can support formation of polariton  gap solitons \cite{Sich2012, Tanese2013,Cerda-Mendez2013,Ostrovskaya2013,Whittaker2018,Zezyulin2018ACS,Zezyulin2020}. 
%
These localized states   can be found using the substitution $\Psi_{1,2}(t, \ell) = e^{-i\mu t} \psi_{1,2}(\ell)$, where stationary wavefunctions $\psi_{1,2}(\ell)$ satisfy zero boundary conditions at $\ell \to \infty$ and $\ell \to -\infty$, and $\mu$ characterizes the chemical potential of the polariton condensate. 
The numerical study indicates that the system supports a variety of solitons which form continuous families, i.e., can be parameterized by the continuous change of the chemical potential $\mu$ within the energy spectrum bandgap.  To describe the found solitons, we introduce the polariton density integral  $N=\int_{-\infty}^\infty (|\psi_1|^2 +  |\psi_2|^2)d\ell$  which characterizes the squared norm of the solution.
In Fig.~\ref{fig:family}(a) we illustrate the family of fundamental (simplest) gap solitons as a dependence $N$ on $\mu$. The soliton family detaches from the left edge of the bandgap, where the soliton norm vanishes: $N \to 0$. In this limit, small-amplitude solitons transform to a linear Bloch wave. As the chemical potential increases towards the right gap edge, the total norm $N$ grows monotonously. To quantify the degree of the soliton localization, we introduce an additional characteristics   $n_{99}$  which amounts to the number of spatial periods where 99\% of quasiparticles are confined. The dependence $n_{99}$ on $\mu$ is also plotted in Fig.~\ref{fig:family}(a). It demonstrates nonmonotonic behavior approaching its minimal values in the center of the gap. In this regime the solitons are most localized, and almost all energy can be trapped in the  segment of waveguide   composed of approximately from five to ten  unit cells.  At the same time, the quantity $n_{99}$ becomes extremely large near the edges of the gap, which means that the corresponding solitons are very broad and relatively poorly localized. Examples of spatial profiles of solitons having different amplitudes and degrees of localization are shown in Fig.~\ref{fig:family}(b). 

It is known that gap solitons   and,  in particular, those  in systems dominated by  repulsive nonlinearities, can be be prone to dynamical instabilities \cite{Louis2003,Efremidis2003,Pelinovsky2004,KIZIN201658}.  In the meantime, using the dynamical simulations, we found that the family of fundamental gap solitons presented in Fig.~\ref{fig:family}(a) contains stable solutions which can robustly preserve the steady shape for the indefinite simulation time (much larger than typical polariton lifetimes), even if the initial profiles are perturbed by a small-amplitude random noise. Example of such stable dynamics is presented in Fig.~\ref{fig:family}(c,d). At the same time, more complex solitons can develop dynamical instabilities which eventually lead to their delocalization. The corresponding example is shown in Fig.~\ref{fig:family}(e,f). 
 

\textit{Conclusion}. In conclusion, we constructed a theory of the propagation of cavity polaritons in narrow quasi-1D waveguides of arbitrary shape and applied it to the case of periodically curved waveguides. We demonstrated that the periodic rotation of an effective in-plane magnetic field produced by TE-TM splitting in linear polarizations leads to the formation of nontrivial band structure. The shape of the bands, the bandgaps and the positions of the band extrema can be tuned by application of an external magnetic field. In the nonlinear regime the system supports formation of dynamically stable gap solitons. 

\begin{acknowledgments}

\textit{Acknowledgements}. The research  was supported by   Priority 2030 Federal Academic Leadership Program. IAS acknowledges support from Icelandic Research Fund (Rannis), project   No. 163082-051.

\end{acknowledgments}


\newpage

\section{Supplemental Material: Derivation of the 1D adiabatic Hamiltonian}

The two-dimensional Hamiltonian of a polariton moving inside a waveguide defined by a confining potential $U(x,y)$ is \cite{Flayac2010}:
\begin{widetext}
	\begin{equation}
	\label{eq:H}
	\hat{H}_{2D} = \left(\begin{array}{cc}
	\displaystyle -\frac{\hbar^2}{2m_{eff}}\left(\frac{\partial^2\ }{\partial x^2}+\frac{\partial^2\ }{\partial y^2}\right)+ \frac{\Delta_z}{2}+U(x,y)&  \beta\left(\frac{\partial }{\partial y}+i\frac{\partial }{\partial x}\right)^2  \\[3mm]
	\beta\left(\frac{\partial }{\partial y}-i\frac{\partial }{\partial x}\right)^2 & \displaystyle   -\frac{\hbar^2}{2m_{eff}}\left(\frac{\partial^2\ }{\partial x^2}+\frac{\partial^2\ }{\partial y^2}\right)- \frac{\Delta_z}{2}+U(x,y)
	\end{array}\right),
	\end{equation}
\end{widetext}
where
\begin{equation}
\beta=\frac{\hbar^2}{4}\left(\frac{1}{m_l}-\frac{1}{m_t}\right).
\end{equation}

Let us introduce in each point of a waveguide local coordinate system with axis $\ell$ directed tangential to it and $n$ normal to it. The elementary lengths $d\ell$ and  $dn$ read:
\begin{eqnarray}
d\ell =\tau_x(\ell)dx+\tau_y(\ell)dy,\\
dn=-\tau_y(\ell)dx+\tau_x(\ell)dy
\end{eqnarray}
where $\tau_{x,y}$ are components of the unit vector tangential to the waveguide at a given point characterized by coordinate $\ell$ along the waveguide.

We can now right down:
\begin{eqnarray}
\frac{\partial\, }{\partial x}=\frac{\partial \ell}{\partial x}\frac{\partial\, }{\partial \ell}+\frac{\partial n}{\partial x}\frac{\partial }{\partial n}=\tau_x\frac{\partial }{\partial \ell}-\tau_y\frac{\partial }{\partial n},\\
\frac{\partial\, }{\partial y}=\frac{\partial \ell}{\partial y}\frac{\partial\, }{\partial \ell}+\frac{\partial n}{\partial y}\frac{\partial\, }{\partial n}=\tau_y\frac{\partial\, }{\partial \ell}+\tau_x\frac{\partial\, }{\partial n},\\
\frac{\partial\, }{\partial y}\pm i\frac{\partial\, }{\partial x}=\pm i\tau_\mp\frac{\partial\, }{\partial \ell}+\tau_\mp\frac{\partial\, }{\partial n},
\end{eqnarray}
where
\begin{equation}
\tau_\pm=\tau_x\pm i\tau_y.
\end{equation}
We thus have:
\begin{eqnarray}
\frac{\partial^2\ }{\partial x^2}+\frac{\partial^2\ }{\partial y^2} =\frac{\partial^2\ }{\partial \ell^2}+\frac{\partial^2\ }{\partial n^2}+\left(\tau_y\frac{\partial\tau_x}{\partial \ell}-\tau_x\frac{\partial\tau_y}{\partial \ell}\right)\frac{\partial}{\partial n},
\end{eqnarray}
where we used that 
\begin{equation}
\tau_x^2+\tau_y^2=1.
\end{equation}
Similarly
\begin{eqnarray}
\left(\frac{\partial}{\partial y}\pm i\frac{\partial}{\partial x}\right)^2=\\
\nonumber=\tau_\mp^2\frac{\partial^2\ }{\partial n^2}-\tau_\mp\frac{\partial}{\partial \ell}\tau_\mp\frac{\partial}{\partial \ell}\pm i\tau_\mp\left(\tau_\mp\frac{\partial}{\partial \ell}+\frac{\partial}{\partial \ell}\tau_\mp\right)\frac{\partial}{\partial n}.
\end{eqnarray}

Let us now suggest that the confining potential locally depends on the transverse coordinate $n$ only, and use adiabatic approximation for the spinor wavefunction $\Psi(x,y)$ representing it as:
\begin{equation}
\Psi(x,y)=\psi(\ell)\phi(n),
\end{equation}
where the part $\psi(\ell)$ describes the propagation of the polaritons along the waveguide, and $\phi(n)$ corresponds to their 1D lateral confinement and can be taken real. This approximation holds if an effective thickness of a waveguide $d$ is much less then its local curvature $R$, which for a parametrically given curve is given by
\begin{equation}
R=\frac{\left[x'(\xi)^2+y'(\xi)^2\right]^{3/2}}{|x'(\xi)y''(\xi)-y'(\xi)x''(\xi)|}.
\end{equation}

Multiplying the Schr\"odinger equation $\hat{H}_{2D}\Psi=E\Psi$ by $\phi(n)$ and integrating by $n$ from $-\infty$ to $+\infty$, one gets for the dynamics of the propagation along the channel the following 1D Schr\"odinger equation:
\begin{equation}
\hat{H}\psi(\ell)=E\psi(\ell),
\end{equation}
where
\begin{widetext}
	\begin{equation}
	\label{eq:H}
	\hat{H} = \left(\begin{array}{cc}
	\displaystyle E_0-\frac{\hbar^2}{2m_{eff}}\frac{d^2\ }{d \ell^2}+ \frac{\Delta_z}{2}& \displaystyle  \Omega_--\beta\tau_-\frac{d}{d \ell}\tau_-\frac{d}{d \ell}  \\[6mm]
	\displaystyle  \Omega_+ -\beta\tau_+\frac{d}{d \ell}\tau_+\frac{d}{d \ell} & \displaystyle E_0-\frac{\hbar^2}{2m_{eff}}\frac{d^2\ }{d \ell^2}- \frac{\Delta_z}{2}
	\end{array}\right),
	\end{equation}
\end{widetext}
and  we  have used that 
\begin{equation}
\int_{-\infty}^{+\infty}\phi(n)\frac{d\phi}{d n}\, dn=0,
\end{equation}
and
\begin{equation}
E_0=\int_{-\infty}^{+\infty}\phi(n)\left(-\frac{\hbar^2}{2m_{eff}}\frac{d^2\ }{d n^2}+U(n)\right)\phi(n)dn
\end{equation}
is the energy of the confinement, and
\begin{equation}
\Omega_\pm=\beta\tau_\pm^2\int_{-\infty}^{+\infty}\phi(n)\frac{d^2\phi}{\partial n^2}dn\approx\frac{\beta}{d^2}\tau_\pm^2=\Omega_0\tau_\pm^2,
\end{equation}
where $d$ is an effective width of the confining channel, and we used Gaussion approximation, $\phi(n)=d\sqrt{\pi}e^{-n^2/(2d^2)}$

Note, that $E_0$ is just a constant, which can be safely dropped. As for the off-diagonal terms $\beta\tau_\pm\frac{d}{d \ell}\tau_\pm\frac{d}{d \ell}
$, one can note, that by the order of magnitude $d/d\ell\sim k$, where $k$ is a wavenumber, describing the propagation of the polaritons along the waveguide. Therefore, for narrow waveguides and small $k$, when $k\ll d^{-1}$, these terms can be neglected as compared to $\Omega_\pm$, and one gets the Hamiltonian (4) of the main text.

\end{document}